\documentclass[amssymb,amsmath,aps,showpacs,twocolumn]{revtex4}

\newcommand{\lsim}{\lesssim}
\newcommand{\gsim}{\gtrsim}

\def\lsim{\mathrel{\raise.3ex\hbox{$<$\kern-.75em\lower1ex\hbox{$\sim$}}}}
\def\gsim{\mathrel{\raise.3ex\hbox{$>$\kern-.75em\lower1ex\hbox{$\sim$}}}}

\def\beq{\begin{equation}}
\def\eeq{\end{equation}}
\def\beqn{\begin{eqnarray}}
\def\eeqn{\end{eqnarray}}
\def\bea{\begin{eqnarray}}
\def\eea{\end{eqnarray}}
\def\be{\begin{equation}}
\def\ee{\end{equation}}
\newcommand{\fslash}[1]{{#1 \kern -0.7em/ \kern 0.1em}}

\begin{document}

\voffset 1.25cm

\title{U-boson at BESIII}
\author{ Shou-hua Zhu}
\affiliation{Institute of Theoretical Physics, School of Physics,
Peking University, Beijing 100871, China}

\date{\today}

\begin{abstract}

The $O$(MeV) spin-1 U-boson has been proposed to mediate the
interaction among electron-positron and $O$(MeV) dark matter, in
order to account for the 511 keV $\gamma$-ray observation by
SPI/INTEGRAL. In this paper the observability of such kind of
U-boson at BESIII is investigated through the processes $e^+e^-
\rightarrow U \gamma$ and $e^+e^- \rightarrow J/\Psi \rightarrow
e^+e^- U$. We find that  BESIII and high luminosity B-factories
have the comparable capacity to detect such U-boson. If U-boson
decays mainly into dark matter, i.e. invisibly, BESIII can measure
the coupling among U-boson and electron-positron $g_{eR}$ (see
text) down to $O(10^{-5})$, and cover large parameter space which
can account for 511 keV $\gamma$-ray observation. On the other
hand, provided that U decays mainly into electron-positron, BESIII
can detect $g_{eR}$ down to $O(10^{-3})$, and it is hard to
explore 511 keV
 $\gamma$-ray measurement allowed parameter space due to the
 irreducible QED backgrounds.

\end{abstract}


\maketitle

\section{Introduction}

The light dark matter, say $1\sim 100\ MeV$ \cite{Boehm:2003bt},
was proposed to account for the SPI/INTEGRAL observation of a 511
keV bright $\gamma$-ray line from the galactic bulge
\cite{511KeV}. The $\gamma$-ray is supposed to be produced via
non-relativistic electron and positron which originate directly
from the dark matter annihilation. Here the dark matter can be
Majorana fermion or scalar particle. Moreover the particle which
is mediating the interactions among dark matter and usual standard
model (SM) matter, in this case electron and positron, can be the
extra gauge boson \cite{Boehm:2003bt}.
 In literature there may be other
alternative mechanism in the next-to-minimal supersymmetric standard
model (NMSSM) \cite{Gunion:2005rw}. The 511 keV $\gamma$ ray arises
also from electron-positron annihilation, but electron and positron
do not directly come from the dark matter annihilation. Instead they
are the decay products of muon pair which arise from the dark matter
$\chi^0$ annihilation via a light pseudo-scalar $a$ with mass $2
m_{\chi^0}\pm 10$ MeV. Here the dark matter is the lightest
neutralino $\chi^0$ with the mass $O$(100-200 MeV)
\cite{Gunion:2005rw}. However there are astrophysical limits which
disfavor such scenario.  Especially the EGRET bounds on
bremstrahlung radiation \cite{Beacom:2004pe} can be translated into
an upper bound to the mass of the dark matter of 20 MeV; and another
bound from galactic positron production \cite{Beacom:2005qv} lowers
the dark matter mass down to 3 MeV or less.

For completeness, in the following we will discuss the general
properties of the new particle which is mediating the interactions
among such low mass dark matter and usual SM fermions. The new
particle is generally denoted as X. In order to be compatible with
the low energy precise measurements, X should have tiny coupling
strength with the SM fermions. If X is scalar or pseudo-scalar
boson, as it usually expected, it will mix with the scalar or
pseudo-scalar which are charged under the SM gauge group. Thus X
tends to couple to the SM fermion with strength which is
proportional to fermion mass, and this is the feature of the Higgs
boson in the SM. Therefore such kind of boson preferably shows up
in heavy fermion sectors. In this sense B-factories have
advantages over charm-factories in detecting such kind of new
physics beyond the SM.  For example the above mentioned light
$\chi^0$ and $a$ can be easily checked in Upsilon decay of $
\Upsilon \rightarrow \gamma a$ at B-factories
\cite{Dermisek:2006py}. However $J/\Psi$ is less sensitive to such
mechanism  and the branching ratio is tiny $Br(J/\Psi \rightarrow
\gamma a) \sim 10^{-7}-10^{-9}$ \cite{Dermisek:2006py}. It should
be noted that such kind of light pseudo-scalar $a$ may have been
observed by HyperCP with mass around 214 MeV \cite{Park:2005ek}.
The light pseudo-scalar $a$ decays mainly into di-muon which
appears very collinear at colliders. Moreover $a$ may play the key
role in search of the SM-like Higgs boson at colliders
\cite{Zhu:2006zv}.

If X is singlet vector or pseudo-vector boson and mixes with the
SM gauge boson, the couplings among X with the SM fermion are {\em
universal} because the SM gauge bosons couple with the SM fermions
universally. Therefore the B-factories and charm-factories have
the same positions to check such kind of new physics beyond the
SM. The key question for both experiments is whether they can
collect enough data samples or not. It should be noted that the
advantages for such {\em mixing scenario between X and SM
counterpart (Higgs boson or gauge boson) } can induce naturally
tiny couplings among X and usual fermions because of the vanishing
small mixing.

Alternatively X can be the gauge boson of the new gauge group and
couple directly to the SM fermions with the tiny couplings.
Obviously, the tiny couplings need to be understood. In this case
not all of the SM fermions are necessarily charged under new gauge
symmetry. Experimentally, the precise measurements for the first
two generations, for example muon and electron anomalous magnetic
moment, have constrained the new gauge symmetry severely, which is
the motivation to search for new physics in the heavy fermion
sector \cite{McElrath:2005bp}. However the charm- and B-factories
have the same opportunities to search for such new physics,
because we have no idea yet to which fermion the new gauge boson
couples preferably.

Recently the observability of the new gauge boson (dubbed as U-boson
in Ref. \cite{Boehm:2003bt,EarlyUBoson}) at low energy linear
colliders is seriously investigated \cite{Borodatchenkova:2005ct}.
The authors concluded that the B-factories and $\Phi$-factories can
cover large parameter regions via the process $e^+ e^- \rightarrow U
\gamma$ in which U decays invisibly or into electron and positron,
depending on whether $m_U < 2 m_\chi$. Here $m_U$ and $m_\chi$
denote the masses of U-boson and dark matter respectively. In this
paper we discuss the possibilities of searching for such kind of
U-boson in the production process of $e^+e^- \rightarrow U \gamma$
and in $e^+e^- \rightarrow J/\Psi \rightarrow e^+e^- U$ at BESIII.

It should be pointed out that the low energy linear colliders with
$\sqrt{s} \ll m_Z$ are ideal places to detect the invisible decay
mode of low mass U-boson. This is the case that $m_U > 2 m_X$ and
U-boson decays preferably into dark matter. The reason is that the
SM backgrounds to the signal processes involve always with the
neutrino, which is highly suppressed at least by a factor of
$O\left(\frac{s}{m_Z^2}\right)$. It should be noted that several
experiments have recently measured invisible decay mode for $\eta,
\eta^\prime$ \cite{Ablikim:2006eg} and $\Upsilon(1S)$
\cite{Tajima:2006nc,Rubin:2006gc}. On the contrary, for the case
of $m_U < 2 m_X$, U-boson will decay mainly into electron and
positron. The SM QED backgrounds for the signal processes are
usually huge as we will see below.

Actually the extra gauge boson has its own right to be
investigated at experiments. In many new physics beyond the SM,
there always exist new gauge bosons. For example in the left-right
symmetric model the gauge group is extended to $SU(2)_L \times
SU(2)_R \times U(1)_{B-L}$, and extra W and Z will appear. If the
extra gauge coupling is not much smaller than that of electro-weak
interaction like the case in left-right symmetric model, the new
physics scale has been pushed to a higher scale, say more than 1
TeV, in order to escape the constraints from current low energy
experiments such as $K_0-\overline{K_0}$ mixing, and the direct
search experiments at LEP and Tevatron etc. On the other hand, the
alternative is still possible, i.e. the new physics scale can be
greatly lowered provided that the new couplings are much smaller
than that of the SM ones. At low energy regime with $Q \ll m_Z$,
it is even possible that the new force is stronger than that of
weak interaction, which is highly suppressed by a factor of
$O\left(\frac{Q^2}{m_Z^2}\right)$. In literature, there are
relatively extensive investigations on the former case. However
the SPI/INTEGRAL measurements may indicate that the latter case
should be taken seriously.

\section{Effective Lagrangian for U-boson and its decay}

Currently the origin of U-boson is not clear. Therefore we start
with the effective interaction lagrangian among
electron-positron-U and $\chi-\chi^*-U$ where $\chi$ denotes
generally dark matter particle. The effective lagrangian can be
written as \cite{Boehm:2003hm}
\begin{eqnarray}
&U-{e^+}-{e^-}&: \gamma_\mu \left( g_{eL} P_L+ g_{eR} P_R \right)
\nonumber \\ & U-{\chi^*}-{\chi}& : g_\chi
\left(p_\chi-p_{\chi^*}\right)\ \ \  \chi={\rm scalar} \nonumber
\\ & &: \gamma_\mu \left( g_{\chi L} P_L+ g_{\chi
R} P_R \right) \ {\rm or} \ g_{\chi}\gamma_\mu  P_R \nonumber \\
&& \chi={\rm Dirac\ or \ Majorana \ fermion}
\end{eqnarray}
with $P_{L,R}=\frac{1}{2}(1\mp \gamma_5)$. As mentioned above,
U-boson may or may not couple to other SM fermions. In this paper
we will conservatively assume U-boson only couples to dark matter
and electron-positron. For the case that U-boson couples to
quarks, the detection of U-boson has been investigated in Ref.
\cite{Fayet:2006sp}. It should be emphasized that the cold dark
matter relic density constrains the the combination of both
couplings while the collider measurements are sensitive usually
only to $g_{eL}$ and $g_{eR}$.

The low energy precise measurements, for example electron $g_e-2$
and electron-neutrino scattering measurements, have severely
constrained the left-handed couplings $g_{eL}$ and $g_\nu$.  In
order to account for the 511 keV $\gamma$-ray observation while
satisfying the low energy constraints, we will assume, as that in
Ref. \cite{Borodatchenkova:2005ct}, (a) $g_\chi \sim O(1)$ and the
couplings among U-electron-positron are much smaller than that of
those of U-dark-matter; (b) $g_{eL}=g_\nu=0$. In Ref.
\cite{Boehm:2003hm} the authors have analyzed the constraints from
$g-2$ measurements in detail and got the corresponding bounds on the
couplings of U-boson with dark matter, as well as with
electron-positron. In their analysis they have assumed
$g_{eL}=g_{eR}$, i.e. vectorial couplings. 

Based on the assumptions above, we will discuss the U-boson decay
modes while varying its mass. The dominant decay modes are $U
\rightarrow e^+e^-$ for $m_U < 2 m_\chi$, i.e. U-boson decay into
dark matter is kinematically forbidden. On the other hand, the
dominant U-boson mode is $U\rightarrow \chi \chi^*$ (invisible
decay) for $m_U > 2 m_\chi$, which is due to the choice that
$g_\chi \gg g_{eL}, g_{eR}$ and $g_\nu$.

\section{General considerations for search of U-boson at BESIII}

BESII has successfully collected $5.8 \times 10^{7}$ $J/\psi$
events. After upgrade to BESIII, luminosity of $e^+e^-$ collision
will reach $10^{33} cm^{-2} s^{-1}$ or equivalent $10^{10}$
$J/\Psi$ events at $\sqrt{s}=3.097$ GeV will be obtained per year
\cite{YFWang}. In our numerical estimations we choose $e^+e^-$
integrated luminosity as $20 fb^{-1}$ or equivalent $4 \times
10^{10}$ $J/\Psi$ events at $\sqrt{s}=3.097$ GeV, which
corresponds to data samples collected within four years.

Because we are interested in the $e^+e^-$ collider with
$\sqrt{s}=3.097$ GeV at $J/\Psi$ mass peak, the amplitude for the
certain final state $f$ can be decomposed into
\begin{eqnarray}
M(e^+e^- \rightarrow f)=M_{direct}+M_{J/\psi}
\end{eqnarray}
where $M_{direct}$ and $M_{J/\psi}$ represent the amplitudes via
$e^+e^- \rightarrow f$ directly (non-resonant production) and
$e^+e^- \rightarrow \gamma^* \rightarrow J/\psi \rightarrow f$,
respectively. In general both $M_{direct}$ and $M_{J/\psi}$ will
contribute to the production rate of the final state $f$. In the
specific case of $f=U+\gamma$, provided that U does not couple to
charm quark, $J/\Psi$ decays into $U+\gamma$ is absent at tree
level. On the contrary, if U couples to charm quark,
 $J/\Psi$ can decay into $U+\gamma$ at tree level. In this paper
we have assumed that U couples only to electron-positron, thus
$M(e^+e^- \rightarrow U\gamma)$ is equal to $M_{direct}$ in the
leading approximation. For the SM backgrounds arising from
$M_{J/\Psi}$ and $M_{direct}$, they can be investigated separately
at $\sqrt{s_{e^+e^-}}= m_{J/\Psi}$ utilizing final states
different angular distributions etc. For simplicity, in our study
we only consider the ones from $M_{direct}$.

For another case with $f=e^+e^-U$, U radiates from
electron-positron legs in the process $e^+e^- \rightarrow e^+e^-$.
Usually this three particles final state will have much smaller
cross section than that of $\gamma U$. However the production
rates can be enhanced in the case (1) of the small angle (large
pseudo-rapidity) region due to the t-channel coulomb singularity;
(2) of $\sqrt{s_{e^+e^-}}$ around the $m_{J/\Psi}$ due to the
s-channel resonance ${J/\Psi}$. Realistically there are
difficulties to utilize the former to detect new physics beyond
the SM. Therefore in this paper we will explore the latter case
and consider the signal and background only in $J/\Psi$ decay for
simplicity.


Throughout the paper, we utilize the package CompHEP
\cite{Pukhov:1999gg} to simulate signal and corresponding
background processes after appropriate modifications of the model
file. As a cross check we have re-produced the results in Ref.
\cite{Borodatchenkova:2005ct}.

\section{Searching for U-boson via
$e^+ e^-\rightarrow U\gamma$ process}

As mentioned above, U-boson associated production with photon
arises from direct $e^+e^-$ annihilation, provided that U-boson
does not couple to charm quark. In the following we will discuss
the search strategies for the two cases (1) U-boson decays mainly
into dark matter because of $m_U > 2 m_\chi$ and appears as
invisible decay mode, and (2) U-boson decays mainly into
electron-positron for $m_U <2 m_\chi$.

\subsection{$U \rightarrow \chi \chi^*$}

The signal process is
\begin{eqnarray}
e^+e^- \rightarrow U \gamma \rightarrow  \gamma + \ missing\
energy.
\end{eqnarray}

The main irreducible SM backgrounds for the signal process are
\begin{eqnarray} e^+e^- \rightarrow \nu \bar \nu \gamma
\end{eqnarray}

\begin{figure}[thb]
\vbox{\kern2.6in\includegraphics{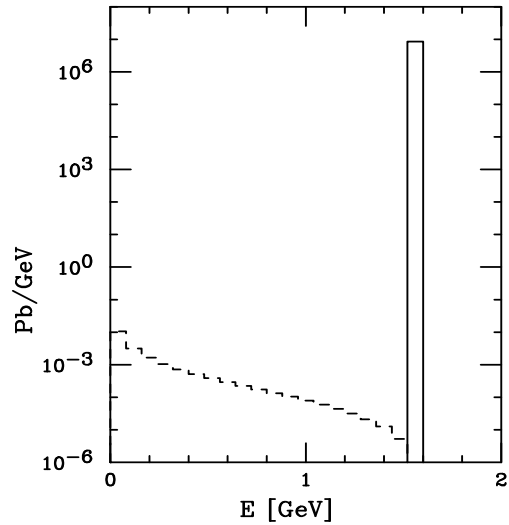}} \caption{  Photon energy distribution of
signal (solid line) and background (dashed line) for $e^+e^-
\rightarrow \gamma U$ with $|\cos\theta_\gamma|<0.99$, $m_U=20$
MeV and $g_{eR}=1$. } \label{fig3}
\end{figure}

In Fig. \ref{fig3} we show the photon energy distribution for
signal
 and background for $e^+e^- \rightarrow
\gamma U$ with $g_{eR}=1$ and $|\cos\theta_\gamma|<0.99$. Here
$\theta_\gamma$ corresponds to the angle among electron beam line
and photon. As shown in the figure, the signal contains one
mono-energetic photon with energy $E=(s-m_U^2)/(2\sqrt{s})\simeq
\sqrt{s}/2$, and background has the continuous photon. Note that
the background is higher-order, i.e. $O(\alpha G_F^2 s)$ process
compared to the signal one. Even with the simple cut, for example
$E_\gamma
>100 MeV$, the background at low energy $e^+e^-$ colliders is
negligibly small.

\begin{figure}[thb]
\vbox{\kern2.6in\includegraphics{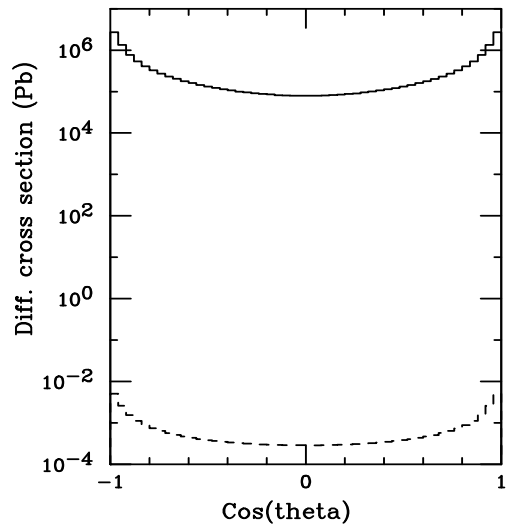}} \caption{ Differential cross section as a
function of $\cos\theta_\gamma$  of signal (solid line) and
background (dashed line)  for $e^+e^- \rightarrow \gamma U$. Here
$g_{eR}=1$, $m_U=20$ MeV and $E_\gamma
>100 MeV$ } \label{fig4}
\end{figure}

In Fig. \ref{fig4} we show the differential cross section as a
function of $\cos\theta_\gamma$ for signal and background with
$g_{eR}=1$.  The shapes of the differential cross section for
signal and backgrounds are similar.
Therefore we impose the cuts as following
\begin{eqnarray}
&& E_\gamma > \sqrt{s}/2- 200 MeV \nonumber \\
&&|\cos(\theta_\gamma)|<0.9.
\end{eqnarray}

\begin{figure}[thb]
\vbox{\kern2.6in\includegraphics{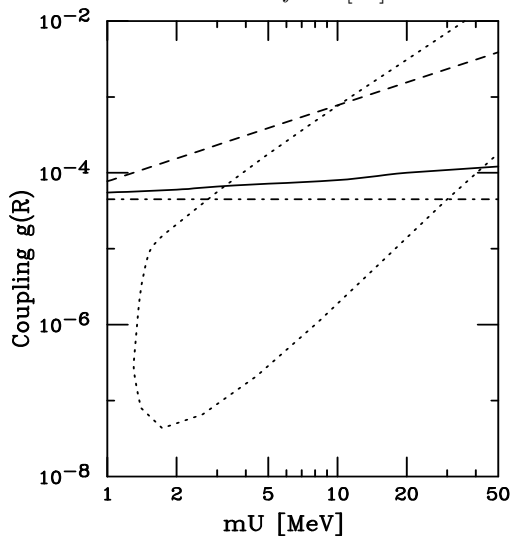}} \caption{ Lower limit of $g_{eR}$ which can
be detected by BESIII with $4\times 10^{10}$ (or equivalent 20
$fb^{-1}$ $e^+e^-$ luminosity) as a function of $m_U$ with $m_U >
2 m_\chi$, i.e. U decays mainly into dark matter. Here solid
(dot-dashed) line represents limit from $J/\Psi \rightarrow e^+e^-
U$ ($e^+e^- \rightarrow U \gamma$). In between the dotted line
represents the parameter space which can account for 511 keV
$\gamma$-ray observation and dashed line indicates the upper bound
from $g_e-2$ measurement with $g_\chi=1$ in the case of complex
scalar dark matter \cite{Borodatchenkova:2005ct}. }\label{fig1}
\end{figure}

In Fig. \ref{fig1} we show the lower limit  of $g_{eR}$
(dot-dashed line) as a function of $m_U$ with $S/\sqrt{B}>5$, in
which $S$ and $B$ represent the number of events for signal and
background respectively. Also shown in Fig. \ref{fig1} are the
parameter space which can account for 511 keV $\gamma$-ray
observation, as well as constraints for $g_{eR}$ from $g_e-2$ with
$g_\chi=1$ in the case of complex scalar dark matter. It is
obvious that BESIII can cover a lot of allowed region which is not
touched by $g_e-2$ measurements. The capacity to detect U-boson at
BESIII is comparable to those of B-factories with $500 fb^{-1}$
\cite{Borodatchenkova:2005ct}.

\subsection{$U \rightarrow e^+e^-$}
\begin{figure}[thb]
\vbox{\kern2.6in\includegraphics{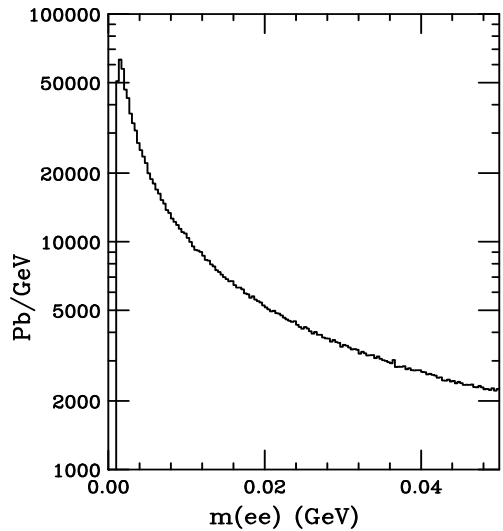}} \caption{  $m_{e^+e^-}$ distribution of the
SM background $e^+ e^-\rightarrow e^+ e^- \gamma$ with second cut
in Eq.\ref{eq6}. }\label{figmee}
\end{figure}

For the decay mode that U-boson decays into electron and positron,
the signal is $e^+ e^-\rightarrow U \gamma \rightarrow e^+ e^-
\gamma$. In signal process $m_{e^+e^-}$ peaks around $m_U$, and
the irreducible background has the $m_{e^+e^-}$ peaks around
$\sqrt{s}$ and several $m_e$ due to t-channel contributions with
soft photon and s-channel contributions, respectively. In Fig.
\ref{figmee} we show the $m_{e^+e^-}$ distribution of background
with $m_{e^+e^-}< 50$ MeV, which is our interested kinematical
region. Obviously the resolution of $m_{e^+e^-}$ is extremely
important to suppress the background. In order to get rid of
background, we impose the cuts as following, similar to that in
Ref. \cite{Borodatchenkova:2005ct},
\begin{eqnarray}
&& m_U-1 MeV < m_{e^+e^-} < m_U+1 MeV \nonumber \\
&&|\cos(\theta_i)|<0.9, \label{eq6}
\end{eqnarray}
where $\theta_i$ (i=$e^+, e^-, \gamma$) corresponds to the angle
among initial electron beam line and final state particle
respectively.  Here the mass spread in the first cut is given by
the BESIII mass resolution, which is valid up to 100 MeV. The
first cut in Eq. \ref{eq6} also implies that the mono-energetic
photon has the energy as $(s-m_{e^+e^-}^2)/(2\sqrt{s})$. The huge
QED background has been suppressed at least two orders of
magnitude via cuts in Eq. \ref{eq6}.

\begin{figure}[thb]
\vbox{\kern2.6in\includegraphics{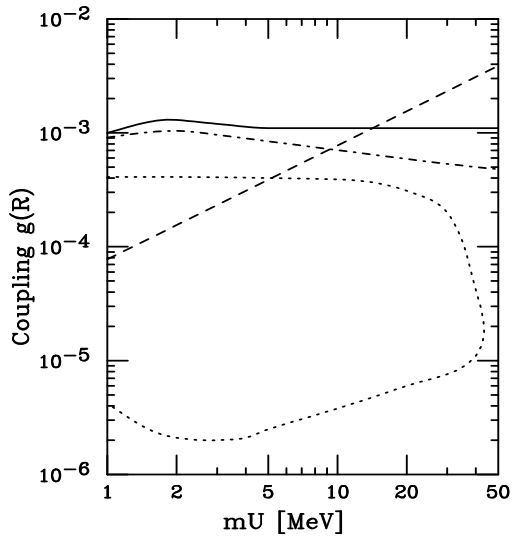}} \caption{  Same with Fig. \ref{fig1}, but
for $m_U < 2 m_\chi$, i.e. U decays mainly into electron-positron.
}\label{fig2}
\end{figure}

Fig. \ref{fig2} shows the lower limit of $g_{eR}$ as a function of
$m_U$ with $S/\sqrt{B}>5$. The conventions in Fig. \ref{fig2} and
Fig. \ref{fig1} are the same, but for $m_U < 2 m_\chi$ i.e.
U-boson decays mainly into electron-positron. we can see that
$g_{eR}$ reach in this channel is rather limited compared to the
case of U-boson invisible decay mode. From Fig. \ref{fig2} we can
see that low energy $e^+e^-$ colliders is incapable of covering
the parameter space which can account for 511 keV $\gamma$-ray
observation.

\section{Searching for U-boson in $J/\psi \rightarrow e^+e^- U$ decay}

The leading mode to produce U-boson in $J/\psi$ decay is $J/\psi
\rightarrow e^+e^- U$ provided that U-boson does not couple to
charm quark, as we have discussed above. Obviously $J/\psi
\rightarrow \mu^+\mu^- U$ is similar to that of electron and
positron if U boson couples to $\mu$. In this paper we focus on
the $e^+e^- U$ case only, but many conclusions can  also  apply
for $\mu^+\mu^- U$.

The significance of $J/\psi \rightarrow e^+e^- U$ mode can be
easily understood if we compare this mode with that of $e^+e^-
\gamma$.
 The branching ratio of $J/\psi \rightarrow e^+e^- \gamma$ is
measured to be $\sim 0.9\%$ with $E_\gamma >100 MeV$
\cite{Armstrong:1996hg}, which is consistent with the SM
predictions. Compared with $Br(J/\psi \rightarrow e^+e^-)\approx
6\%$, the three body decay is not severely suppressed. Thus for
the light MeV U-boson, it is not strange that huge amount of
$J/\psi$ can be used to detect the coupling among U-boson and
electron-positron via three body decay $J/\psi \rightarrow e^+e^-
U$.

\subsection{$U \rightarrow \chi \chi^*$}

For the U-boson invisible decay mode, the final states of signal
will be $e^+e^-$ plus missing energy. The main SM irreducible
background is $J/\Psi \rightarrow e^+e^- \nu_e \bar \nu_e$. For
the signal, the branching ratio $Br(J/\Psi \rightarrow e^+e^-
U)/Br(J/\Psi \rightarrow e^+e^-) \approx 0.46 g_{eR}^2$ and $1.1
g_{eR}^2 $ with $m_U=20$ and $2$ MeV respectively. For the SM
background, $Br(J/\Psi \rightarrow e^+e^- \nu_e \bar
\nu_e)/Br(J/\Psi \rightarrow e^+e^-) \sim 4 \times 10^{-14}$,
which is highly suppressed as we have emphasized previously. Thus
the SM background can be safely neglected.  In order to gauge the
capacity how precisely the U-electron-positron coupling can be
measured, we require at least 10 signal events to be observed. In
Fig. \ref{fig1}, we show the lower limit of $g_{eR}$ (solid line)
as a function of $m_U$. Provided $4 \times 10^{10}$ $J/\Psi$
events at BESIII and $Br(J/\Psi \rightarrow e^+e^-)=0.06$, the
lower limit of $g_{eR}$ is $\approx 1 \times 10^{-4}$ and  $6
\times 10^{-5}$ for $m_U=20$ and 2 MeV respectively. For the U
invisible decay mode, the capacity to detect $g_{eR}$ is
comparable in the processes $e^+e^- \rightarrow U \gamma$ and
$J/\Psi \rightarrow e^+e^- U$.

\subsection{$U \rightarrow e^+e^-$}

For the U-boson decays mainly into electron-positron, the final
states of signal will be $e^+e^-e^+e^-$ and SM irreducible QED
background is large. In order to suppress the background, we
require that the invariant mass for one of the four possible
electron-positron combinations should be within [$m_U$-1 MeV,
$m_U$+1 MeV]. This requirement will suppress the background two
orders of magnitude. The $5\sigma$ limit of $g_{eR}$ is shown as
solid line in Fig. \ref{fig2}. From the figure we can see that
$g_{eR}$ reach in this channel is also limited due to the huge QED
background. Through $J/\Psi \rightarrow e^+e^- U$ channel with
$U\rightarrow e^+e^-$, it is difficulty in covering the parameter
space which can account for 511 keV $\gamma$-ray observation.

\section{Discussions and conclusions}

In this paper we have explored the observability of U-boson at
BESIII through the processes $e^+e^- \rightarrow U \gamma$ and
$J/\Psi \rightarrow e^+e^- U$. Both processes can be utilized to
search for U-boson in an efficient way. We find that BESIII and high
luminosity B-factories have the comparable capacity to detect
U-boson. Especially, provided that U-boson decays mainly into dark
matter i.e. invisibly, the BESIII can measure the coupling among
U-boson and electron-positron $g_{eR}$ down to $O(10^{-5})$, and
cover large region of parameter space which can account for the 511
keV $\gamma$-ray observation by SPI/INTEGRAL \cite{511KeV}. However,
provided that U-boson can't decay into a pair of dark matter
particles due to the kinematical reason ($m_U < 2 m_\chi$), instead
it decays mainly into electron-positron, BESIII can detect $g_{eR}$
down to $O(10^{-3})$. In this case it is hard to explore 511 keV
$\gamma$-ray measurement allowed parameter space due to the huge
irreducible QED backgrounds.

In this paper we focus on the data sample at
$\sqrt{s_{e^+e^-}}=3.097$ GeV only. In fact, for the $U+\gamma$
final states the data sample within $\sqrt{s_{e^+e^-}}= 2-5$ GeV
can also be utilized. Obviously more data implies that the lower
$g_{eR}$ can be reached. Moreover we has assumed that U does not
couple to charm quark, thus $J/\Psi\rightarrow U \gamma$ is absent
at tree level. Once this assumption is removed, $J/\Psi\rightarrow
U \gamma$ will be one important channel to search for new U-boson
\cite{Fayet:2006sp}.

Light dark matter can be studied at low energy linear colliders, in
fact it can be checked independently from astrophysical
observations, as suggested by Ref. \cite{Hooper:2003sh}. Last but
not the least, we would like to emphasize that the low energy linear
colliders are the ideal places to search for the invisible decay
mode in the new physics beyond the SM. The SM background for such
signal will always involve neutrino in the final states, which is
highly suppressed at least by a factor of $O(s/m_Z^2)$.

\section{Acknowledgements}

The author thanks Profs. H.B. Li and Y.F. Wang for the discussions
on BESIII experiments, and Profs. K.T. Chao, C. Liu and S.L. Zhu
for the stimulating discussions. This work was supported in part
by the Natural Sciences Foundation of China under grant No.
90403004 and 10635030, the trans-century fund and the key grant
project (under No. 305001) of Chinese Ministry of Education.

\end{document}